\documentclass[conference]{IEEEtran}
\IEEEoverridecommandlockouts

\usepackage{cite}
\usepackage{comment}
\usepackage{amsmath,amssymb,amsfonts}
\usepackage{algorithmic}
\usepackage{graphicx}
\usepackage{textcomp}
\usepackage{tabularx}
\usepackage{booktabs}
\usepackage[table,xcdraw]{xcolor}
\usepackage{glossaries}
\usepackage{soul}
\usepackage{svg}
\makeglossaries
\newacronym{gprs}{GPRS}{General Packet Radio Services}
\newacronym{egprs}{EGPRS}{Enhanced Data Rates for GSM Evolution}
\newacronym{iot}{IoT}{Internet of Things}
\newacronym{nbiot}{NB-IoT}{Narrowband IoT}
\newacronym{lte}{LTE}{Long Term Evolution}
\newacronym[plural=LPWANs,firstplural=Low-Power Wide-Area Networks (LPWANs)]{lpwan}{LPWAN}{Low-Power Wide-Area Network}
\newacronym{lpwa}{LPWA}{Low-Power Wide-Area}
\newacronym{lorawan}{LoRaWAN}{Long Range Wide Area Network}
\newacronym{lora}{LoRa}{Long Range}
\newacronym{nb}{NB}{Narrowband}
\newacronym{m2m}{M2M}{Machine to Machine}
\newacronym{cbm}{CBM}{Condition Based Maintenance}
\newacronym{ai}{AI}{Artificial Intelligence}
\newacronym{edrx}{eDRX}{Extended Discontinuous Reception Mode}
\newacronym{drx}{DRX}{Discontinuous Reception Mode}
\newacronym{psm}{PSM}{Power Saving Mode}
\newacronym{gps}{GPS}{Global Positioning System}
\newacronym{ptw}{PTW}{Paging Time Window}
\newacronym{tau}{TAU}{Tracking Area Update}
\newacronym{mtc}{MTC}{Machine-Type Communication}
\newacronym{ue}{UE}{User Equipment}
\newacronym{cdrx}{CDRX}{Connected Mode DRX}
\newacronym{3gpp}{3GPP}{3rd Generation Partnership Project}
\newacronym{gsm}{GSM}{Global System for Mobile Communications}
\newacronym{rf}{RF}{Radio Frequency}
\newacronym{prbs}{PRBs}{Physical Resource Blocks}
\newacronym{mcl}{MCL}{Maximum Coupling Loss}
\newacronym{ce}{CE}{Coverage Enhancement}
\newacronym{rsrp}{RSRP}{Reference Signals Received Power}
\newacronym{toa}{ToA}{Time on Air}
\newacronym{rrc}{RRC}{Radio Resource Connection}
\newacronym{rtc}{RTC}{Real Time Clock}
\newacronym{css}{CSS}{Chirp Spread Spectrum}
\newacronym{ism}{ISM}{Industrial, Scientific and Medical}
\newacronym{mmtc}{MMTC}{Massive Machine Type Communication}
\newacronym{sf}{SF}{Spreading Factor}
\newacronym{cr}{CR}{Coding Rate}
\newacronym{bw}{BW}{Bandwidth}
\newacronym{adr}{ADR}{Adaptive Data Rate}
\newacronym{abp}{ABP}{Authentication By Personalisation}
\newacronym{otaa}{OTAA}{Over The Air Authentication}
\newacronym{ofdm}{OFDM}{Orthogonal Frequency Division Multiplexing}
\newacronym{scfdma}{SCFDMA}{Single-Carrier Frequency Division Multiple Access}
\newacronym{udp}{UDP}{User Datagram Protocol}
\newacronym{qos}{QoS}{Quality of Service}
\newacronym{ra}{RA}{Random Access}
\newacronym{rar}{RAR}{Random Access Response}
\newacronym{snr}{SNR}{Signal-to-Noise Ratio}
\newacronym{ota}{OTA}{Over The Air}
\newacronym{ttm}{TTM}{Time To Market}
\newacronym{cots}{COTS}{commercial off-the-shelf}
\newacronym{mac}{MAC}{Medium Access Control}
\newacronym{multi-rat}{Multi-RAT}{Multiple Radio Access Technology}
\newacronym{rat}{RAT}{Radio Access Technology}
\newacronym{pcb}{PCB}{Printed Circuit Board}
\newacronym[plural=WSNs,firstplural=Wireless Sensor Networks (WSNs)]{wsn}{WSN}{Wireless Sensor Network}
\newacronym{uav}{UAV}{Unmanned Aerial Vehicle}
\newacronym{soc}{SoC}{State of Charge}
\newacronym{wpt}{WPT}{Wireless Power Transfer}
\newacronym{ipt}{IPT}{Inductive Power Transfer}
\newacronym{cpt}{CPT}{Capacitive Power Transfer}
\newacronym{rfeh}{RFEH}{Radio Frequency Energy Harvesting}
\newacronym{em}{EM}{Electromagnetic}
\newacronym{eirp}{EIRP}{Effective Isotropic Radiated Power}
\newacronym{pa}{PA}{power amplifier}
\newacronym{gnss}{GNSS}{Global Navigation Satellite System}
\newacronym[plural=WRSNs,firstplural=Wireless Rechargeable Sensor Networks (WRSNs)]{wrsn}{WRSN}{Wireless Rechargeable Sensor Network}
\newacronym{mcu}{MCU}{Microcontroller Unit}
\newacronym{ipy}{IPY}{Interventions per Year}
\newacronym{smps}{SMPS}{Switched Mode Power Supply}
\newacronym{ldo}{LDO}{Low-dropout voltage regulator}
\newacronym{cv}{CV}{Constant Voltage}
\newacronym{cc}{CC}{Constant Current}
\newacronym{rtk}{RTK}{Real Time Kinematics}
\newacronym{lto}{LTO}{Lithium Titanate Oxide}
\newacronym{lco}{LCO}{Lithium Cobalt Oxide}
\newacronym{lic}{LIC}{Lithium-ion Capacitor}
\newacronym{tdoa}{TDoA}{Time Difference of Arrival}
\newacronym{aoa}{AoA}{Angle of Arrival}
\newacronym{rssi}{RSSI}{Received Signal Strength Indicator}
\usepackage[binary-units=true,detect-all = true]{siunitx}
\usepackage[numbers,sort&compress]{natbib}
\usepackage{pgfplots}
\usepackage{caption}
\usepackage{subcaption}
\pgfplotsset{width=0.5\textwidth, height=7cm, compat=1.9}
\usepackage{url}
\def\BibTeX{{\rm B\kern-.05em{\sc i\kern-.025em b}\kern-.08emT\kern-.1667em\lower.7ex\hbox{E}\kern-.125emX}}
\usepackage[inline]{enumitem}
\usepackage{slashbox}
\usepackage{tikz}

\usepackage[utf8]{inputenc}
\usepackage{pgfplots}
\DeclareUnicodeCharacter{2212}{−}
\usepgfplotslibrary{groupplots,dateplot}
\usetikzlibrary{patterns,shapes.arrows}
\pgfplotsset{compat=newest}

\newcounter{liesbet}

\newcounter{guus}
\newcommand{\guus}[1]{{\color{olive}[GUUS \theguus: #1]\stepcounter{guus}}}
\newcounter{jarne}
\newcommand{\jarne}[1]{{\color{violet}[JARNE \thejarne: #1]\stepcounter{jarne}}}
\newcounter{gilles}
\newcommand{\gilles}[2][]{{\color{orange}[GILLES \thegilles: \ifthenelse{\equal{#1}{L}}{$\leftarrow$ }{} #2 \ifthenelse{\equal{#1}{R}}{ $\rightarrow$}{}]\stepcounter{gilles}}}
\newcounter{lieven}

\newcommand{\eg}[0]{e.g., }

\begin{document}

\title{Aerial Energy Provisioning for Massive Energy-Constrained IoT by UAVs}

\author{
    \IEEEauthorblockN{Jarne Van Mulders, Guus Leenders, Gilles Callebaut, Lieven De Strycker and Liesbet Van der Perre}
    \IEEEauthorblockA{
        KU Leuven, ESAT-WaveCore, Ghent Technology Campus\\
                        B-9000 Ghent, Belgium                                                               \\
                        name.surname@kuleuven.be                                                        }
}
\maketitle

\begin{abstract}
Autonomy of devices is a major challenge in many \gls{iot} applications, in particular when the nodes are deployed remotely or difficult to assess places. In this paper we present an approach to provide energy to these devices by \glspl{uav}. Therefore, the two major challenges, finding and charging the node are presented. We propose a model to give the energy constrained node an unlimited autonomy by taken the \gls{wpt} link and battery capacity into account. Selecting the most suitable battery technology allows a reduction in battery capacity and waste. Moreover, an upgrade of existing \gls{iot} nodes is feasible with a limited impact on the design and form factor.
\end{abstract}


\IEEEpeerreviewmaketitle


\section{Introduction}
\gls{iot} applications often require vast \glspl{wsn} to be deployed in remote and inaccessible locations. Sensor nodes can be deployed on agriculture fields, managing and optimizing the irrigation of the crops~\cite{khoa2019smart}. To accurately monitor the health and growth of trees, recent efforts include the continuous monitoring of stem sap flow \cite{thoen2019deployable,valentini2019new}. Most devices in a \gls{wsn} are typically battery powered. To be economically viable, manual maintenance possibilities (\eg replacing batteries) are limited. As such, energy efficiency and longevity of the \gls{iot} node are of paramount importance. 
To improve the autonomy of an \gls{iot} node, several methods have been thoroughly discussed  in literature. Reducing the battery drainage of a node is often the first approach~\cite{callebaut2021art}. The overall energy consumption can be reduced by making use of energy optimization techniques such as choosing energy-efficient communication, introducing stand-by and sleep phases~\cite{callebaut2021art}. However, some accurate sensors and actuators (\eg motorized valves in irrigation systems or sensors requiring heaters) inherently use larger amounts of energy. In such devices, the energy reducing possibilities are limited.  
Another straightforward approach is using a battery with a larger capacity. This improves the autonomy, at the expense of node size, cost, and ecological footprint. 

\begin{figure}
    \centering
    \includegraphics[width=0.4\textwidth]{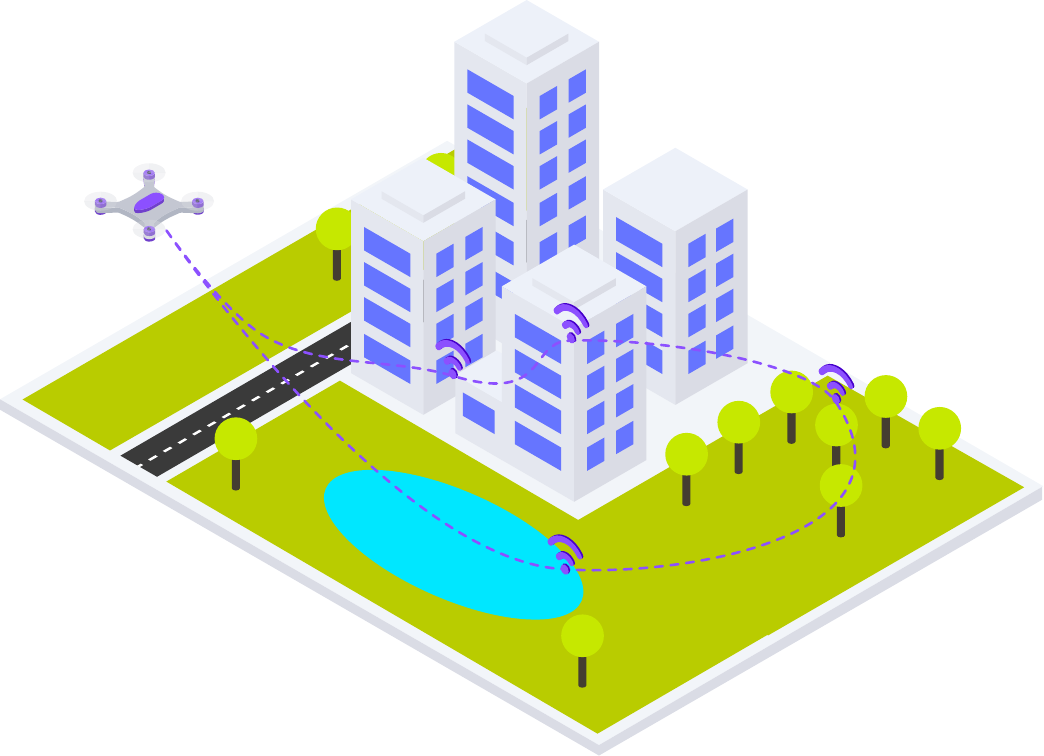}
    \caption{\Gls{uav} navigating to the \gls{iot} nodes. When the \gls{uav} is near the \gls{iot} device, the \gls{uav} initiates the \acrlong{wpt}, charging to the node's battery.}
    \label{fig:intro}
\end{figure}

When external ambient energy sources are available in the use case at hand, the battery can be recharged by utilizing energy harvesting techniques. Most common techniques are harvesting from solar (energy). 
However, these energy sources tend to be of lower energy density \cite{harb2011energy} and are falling short for \gls{iot} use cases with higher power sensors or actuators. In other use cases, no or little energy harvesting possibilities exist, or the extra cost and size of the energy harvesting solution for each individual node is not viable. To combat these battery autonomy issues, we propose a \gls{wsn} charging method using \glspl{uav} and short range wireless power: \glspl{wrsn}~\cite{yang2015wireless}. In this concept, each \gls{iot} node is equipped with a wireless charging mechanism, allowing the \gls{uav} to fly in close proximity and charge the \gls{iot} node. This concept is illustrated in Fig.~\ref{fig:intro}.

The concept of charging energy-constrained \gls{iot} devices, has been proposed in literature~\cite{yao2019energy}. Recent studies explore this concept from the \gls{uav} point-of-view: comparing various scheduling algorithms to optimize the routing path of the \gls{uav}~\cite{wei2019energy, su2020uav}. Yet, in depth research on how an \gls{iot} charging system using \glspl{uav} can be established from a technical node point-of-view, is lacking.


{\textbf{\textit{Contributions}-- }} The main contribution of this work is the development of the technological concepts to achieve a theoretical infinite lifespan of an \gls{iot} node using a \gls{uav}-enabled aerial charging method. 
We clarify and quantify the specific requirements imposed by the \gls{uav}-based charging approach, in particular concerning charging time. We further elaborate on the theoretical limitations and practical restrictions and solutions to achieve sufficiently efficient energy transfer, localization and alignment. Candidate technologies for the actual power transfer are assessed and compared. We present models quantifying the energy transfer, and based on those validate that the selected technologies allow upgrading existing \gls{lpwan} nodes with limited impact on design and form factor.  

This paper is organized as follows. First, we elaborate on \gls{uav} aerial charging concept and explore operating principles and restrictions. Secondly, we discuss technological options and selection for
\begin{enumerate*}[label=(\arabic*)]
    \item charging the \gls{iot} node by \gls{uav}, and
    \item localizing the \gls{iot} node for accurate alignment to start energy transfer.
\end{enumerate*}
Thirdly, we model energy storage and charging systems, resulting in a \gls{uav}-based provisioning profile. Fourthly, we apply our findings on two use cases. Finally, we summarize our main findings. 

\textit{We consider an \gls{iot} node with a battery capacity between \SI{1}{\kilo\joule} and \SI{10}{\kilo\joule} that needs to be charged within a limited time of maximum \SI{5}{\minute}.}

\section{Concept and Operating Principle}\label{concept}
Most common battery-powered \gls{iot} nodes are typically restricted to an autonomy of a few years at best~\cite{thoen2019deployable, gawali2019energy, ashraf2015introducing}. 
We propose to extend the battery time of \gls{iot} nodes to a virtually limitless autonomy using a \gls{uav} aerial recharging concept (Fig.~\ref{fig:intro}). \gls{iot} use cases often deploy a swarm of \gls{iot} nodes for monitoring purposes (\eg environmental monitoring)~\cite{dafflon2021distributed}. Each \gls{iot} node is connected and can send measurements to a central server using a wireless connection (\eg \gls{lpwan} connectivity). To monitor the charge level of the node's battery, the \gls{soc} is a key metric. The \gls{soc} can be acted upon by sending \gls{soc} messages, when the \gls{soc} falls below a certain threshold or by regularly including the \gls{soc} parameter in the node's communication. The central server keeps an overview and predicts the \gls{soc} of each \gls{iot} node. Acting upon this information, it can send a \gls{uav}, equipped with the appropriate charging circuitry, to nodes that are (almost) running out of battery. By being able to predict \gls{soc} information, a more efficient route can be mapped out ~\cite{liu2020wireless, su2020uav}. 
The \gls{uav} responsible for charging a certain \gls{iot} node, faces two main tasks. First, navigating to the node and aligning itself with respect to the node. Second, creating the energy transfer link and start charging the node's battery. These two distinct stages are studied more in depth in the remainder of this work.

By adopting this \gls{uav} recharging concept, each node will vastly prolong its autonomy. This results in several benefits. Firstly, as the lifespan of an \gls{iot} node prolongs, less manual interventions will be required and less devices will need to be replaced. Secondly, \gls{iot} nodes require lower-capacity batteries: reducing the cost and the environmental footprint. Finally, regular \gls{uav} visits can enable proactive maintenance, both by enabling larger software updates and hardware swaps. 

\section{Selection of key technologies for aerial recharging}
Aerial recharging of \gls{iot} nodes raises significant challenges in terms of both the charging itself, which needs to be in a short time, and the localization of the node by the \gls{uav}. 
In the following, we elaborate on the potential technological options for both tasks, and motivate our selection.

\subsection{Task~I: Charging the \gls{iot} node}
\label{charging}
The \gls{iot} node needs a continuous source of energy.
We propose to satisfy the energy need by frequently provisioning energy through \glspl{uav}. As the latter remains in the air during the process, the charging speed is of paramount importance and poses a challenge for the  \gls{uav}. The speed of charging is predominantly determined by the energy storage and the charging method.

\subsubsection{Energy Storage}
Non-rechargeable batteries, commonly used in place-and-forget \gls{iot} nodes \cite{sather2017battery}, have high volumetric and weight energy densities, making them attractive options for energy-constrained \gls{iot} nodes. Moreover, their self-discharge energy is relatively low compared to rechargeable batteries. Clearly, in our proposed system, a technological shift towards rechargeable-only solutions is needed. The selected battery should have a high recharge cycle count and a high shelf life, to ensure a long lifetime, and a low internal resistance to enable fast charging. A \gls{lto} battery, for instance, satisfies these requirements~\cite{han2014cycle}.






\subsubsection{Charging}
To charge the \gls{iot} node, the \gls{uav} should transfer energy to the node. Energy can be transferred either through physical contacts or wireless transmission. 
Physical contacts are exposed to the elements causing \eg corrosion. Furthermore, the \gls{uav} must be more precisely aligned than \acrlong{wpt}. Hence, wireless transmission is preferred, as this does not require exposing circuitry outside the node's enclosure and allows for a less accurate alignment of the \gls{uav}.  
Several candidates for \gls{wpt} are considered in this comparative study: 
\begin{enumerate*}[label=(\arabic*)]
  \item \acrlong{ipt},
  \item \acrlong{cpt},
  \item \acrlong{rfeh}.
\end{enumerate*}
In what follows, we provide a brief discussion on the appropriateness of these options for \gls{uav}-based charging.



\begin{figure} 
     \centering
     \begin{subfigure}[b]{0.24\textwidth}
         \centering
         \includegraphics[width=\textwidth]{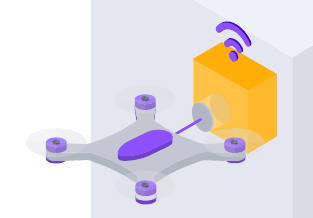}
         \caption{}
         \label{fig:wpt-inductive}
     \end{subfigure}
     \hfill
     \begin{subfigure}[b]{0.24\textwidth}
         \centering
         \includegraphics[width=\textwidth]{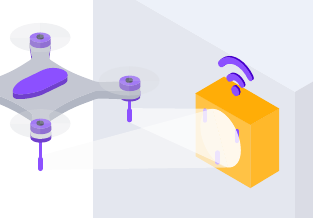}
         \caption{}
         \label{fig:wpt-rf}
     \end{subfigure}
        \caption{Illustration of \gls{uav}-based energy provisioning through \gls{ipt} or \gls{cpt}~(a) or RF~(b)}
        \label{fig:wpt}
\end{figure}

\textbf{\Acrfull{ipt}.}
An inductive link is generated by means of magnetic coupling between two coils, one at the transmitting and one at the receiving side of the energy transfer set-up and is called \gls{ipt}. This is illustrated in Fig.~\ref{fig:wpt}. The operation is  similar to a transformer, which is built with a primary and secondary coil mostly winded around a ferrite or laminated steel core. 
In the aerial charging case, there is air between the two coils. Note that due to the low permeability of air, the self inductance and coupling factor $k$ between the two coils is low in comparison to set-ups that can rely on magnetically conductive materials. 
The efficiency can be improved by using series or parallel connected capacitors on both \gls{wpt} coils to create resonance circuits. This reduces the leakage inductance created by the lower coupling factor over the air. 
More than \SI{1}{\kilo\watt} is achievable through \gls{ipt}. 
Nevertheless, the efficiency strongly depends on the coil alignment, the transfer distance, the size of the coils, and the magnetic losses~\cite{van2009inductive}. 
For the aerial charging approach, provided the aforementioned challenges in efficiency are carefully designed for, \gls{ipt} is an interesting candidate technology.

\textbf{\Acrfull{cpt}.}
Instead of using an magnetic field, an electric field can deliver energy wirelessly by means of capacitors, 
called a \gls{cpt} system. When bringing transmitter and receiver close to each other, two coupling capacitors are effectively created. Both transmitter and receiver consist of two metal plates, which create a capacitive coupled system during the energy transfer. Different structures like the two-plate, four-plate parallel, four-plate stacked structures exist in combination with several circuit topologies~\cite{lu2017review}. Typical systems consist of an inverter, two compensation networks, a capacitive coupler and a rectifier. Due to low capacitance between the plates, resonance is used to produce high voltages, resulting in sufficient electric fields for the energy transfer. 
The switching frequency and plate voltage can be increased to exchange energy over higher distances. Similar to \gls{ipt}, literature shows that power delivery up to more than \SI{1}{kW} is achievable with \gls{cpt}~\cite{lu2017review}. 

\textbf{\Acrfull{rfeh}.}
Energy transmission by means of \gls{em} waves, called \gls{rfeh}, is based on a \gls{pa} with transmitter and a receiver dipole or directive antenna with a matching circuit to charge the device. 
The available energy on the receiver depends on the distance, frequency, transmitted power and gain of the antennas. 
An easy accessible solution is to use the \gls{ism} band, for which the transmitted power and duty cycle are regulated following the corresponding \gls{ism} band regulations
~\cite{etsi2012electromagnetic}. 
\Gls{em}-waves based transfer is appealing as it does not rely on nearby coupling. However, it is important to note that the waves are affected by the path loss, which in free space increases with at least the square of the distance. The latter is not the case for \gls{ipt}/\gls{cpt}. 
We hereby provide a basic case study to illustrate what can (not) be expected from RF-based energy harvesting. 
Consider two antennas at a distance of \SI{30}{cm} and a transmit power of \SI{27}{dBm}. Using the near field transmission formula from~\cite{schantz2013simple}, the received power would be approximately \SI{10}{dBm}. 
Hence, it would take around 100~seconds to transfer 1 joule of energy. 

\textbf{Evaluation of \gls{ipt}, \gls{cpt} and \gls{rfeh}.} In view of technology selection, we here compare the required power, size, proximity and efficiency of the above energy transfer solutions. 
We consider a \textit{power} requirement of \SI{1}{\kilo\joule} to \SI{10}{\kilo\joule} transferred in \SI{5}{\minute}, yielding a \gls{wpt} link of \SI{3}{\watt} to \SI{33}{\watt}. \Gls{ipt} and \gls{cpt} are both suitable technologies to deliver this amount of power. A \gls{rf} link is not able to transfer sufficient energy (in accordance to the regulations of the \gls{ism} band) within the time constraint. Moreover, the \textit{efficiency} for an \gls{rf} system is low in comparison to a \gls{cpt} or \gls{ipt} system, the latter two achieving efficiencies up to \SI{90}{\percent}~\cite{lu2017review}. The \textit{transceiver size} 
is relatively large for \gls{ipt} or \gls{cpt} systems compared to \gls{rf} systems (\eg a dipole antenna), due to the need of coils and capacitor plates. \textit{Alignment} requirements also vary widely among \gls{wpt} technologies. \Gls{rf} features non-critical alignment, aligning \gls{cpt} systems is more critical. In particular, when using a four plate parallel structure, all four plates must be aligned. A typical \gls{ipt} system requires two coils to be aligned within certain margins. For instance, when using the P9221 IC~\cite{P9221R3}, the power transfer efficiency is lowered from \SI{85}{\percent} to \SI{70}{\percent} by a misplacement of \SI{12}{\milli\meter} for a 32 by \SI{42}{\milli\meter} coil. 

In conclusion, we select \Gls{ipt} as the best fit technology for aerial charging, considering the above discussed options, current state of technology, and typical constraints and energy needs of \gls{iot} nodes. 

\subsection{Task~II: Finding the \gls{iot} node}
Prior to charging the nodes, a \gls{uav} needs to find the location of the node and align itself in order to provide contactless energy transfer. We here discuss the technological options to achieve this, and motivate the benefits and shortcomings.
We assume that the \gls{uav} can locate itself with an on-board-\gls{gnss} system and can fly automatically to get near to the depleted node. This is a fair assumption, given that professional \glspl{uav} are by default equipped with \gls{gnss}.
This technology provides a first coarse positioning, which needs to be followed by a precise positioning and alignment step to enable efficient coupled energy transfer. We briefly discuss the technological options and selection for both steps here below.    


\subsubsection{Coarse positioning}
During installation, the operator can enter the geographic coordinates of the node. Alternatively, the location is determined via geolocation or an on-board \gls{gnss}. 

\textbf{Geolocalization.} An \gls{iot} node can position itself using the built-in radio: by using the localization option of many  wireless \gls{iot} (\eg Wi-Fi, Bluetooth, \gls{lpwan}), the position of the \gls{iot} node can be determined. The location is calculated in post-processing, based on the the regular data communication packets (by applying techniques such as \gls{rssi} based localization and trilateration). Using geolocalization proves to be not feasible for \gls{uav} aerial recharging as the currently obtained accuracy is insufficient. For \gls{lpwan} technologies, for example, the best obtained accuracy is \SI{100}{\meter}~\cite{janssen2018outdoor, fargas2017gps}. 


\textbf{\acrlong{gnss}.} 
By adding the required hardware for \gls{gnss} reception, more accurate positioning can be achieved. 
\gls{gps} receivers, for example,  can achieve a localization accuracy of up to \SI{2}{\meter} in open areas and \SI{5}{\meter} in forested landscapes~\cite{ucar2014dynamic}. To achieve an even higher accuracy, e.g. the GPS \gls{rtk} technique can be used, improving accuracy to up to \SI{1}{\centi\meter}. 
With an extra communication link, the location error from a fixed predetermined base station, is sent to the mobile \gls{gps} \gls{rtk} receiver~\cite{cordesses2000combine}. Because of the high accuracy that can be obtained, we consider \gls{gps} \gls{rtk} technology the prime candidate for localization of the \gls{uav}~\cite{stempfhuber2011precise}. 
\subsubsection{Precise positioning and alignment}
As the accurate location of the \gls{iot} node cannot be guaranteed,
a fine-grained localization system should be foreseen to bridge the last meter to centimeters alignment. Especially under bridges, in forests, etc., the \gls{uav} can not rely on the \gls{gnss} system. Two types of implementations for the alignment can be considered, and categorized as either a passive or an active system. They are distinguished from each other by the need for extra components on the \gls{iot} node.


Existing deployments could benefit from a \textbf{passive} system as no additional hardware is required at the node side. Moreover, no extra components could drain out the energy constrained node. Consequently, the finding of the node is determined by the \gls{uav}, without the need for feedback from the node.
An exemplary passive system works on the basis of a camera placed on the \gls{uav}, which is able to detect objects. Object recognition with computer vision requires compute resources and algorithms~\cite{druzhkov2011new} to process the incoming images of the camera. 
Additional sensors can keep the drone at a safe distance from obstacles and assist the alignment. A disadvantage of the camera-based system is its limited operating range when the distance between \gls{uav} and node is large, in combination with the required line-of-sight. 

In contrast, an \textbf{active} solution requires a sensor, actuator or transceiver on both the \gls{uav} and the node, thus supplementary hardware that should be powered by the node. Besides the extra cost, the node can become unreachable if the battery is completely depleted. 
Several technologies have been proposed for precise nearby positioning, based on RF signals or sound sources, and using for example \gls{tdoa} or \gls{aoa} techniques~\cite{8692423}. 
The additional energy consumed by an active solution, can also be tackled by using energy harvesting. \gls{rfeh} on the node side can power a small circuit with a sensor to locate the node. Hence, the node remains detectable, even when the battery is empty. 
For example, through a hybrid  RF-acoustic system~\cite{cox2020energy} having multiple speakers at the \gls{uav} side and a microphone on the node side, the node's position can be determined. The microphone is extended with an \gls{rf} energy harvester, able to receive enough power to send back the collected acoustic data via backscattering. Further alignment improvements in for example \gls{ipt}, could rely on monitoring the link efficiency or on introducing multiple alignment coils~\cite{P9221R3}.
\label{gains}

\section{Modelling the Energy Storage and Provisioning Profile}



The number of \gls{uav} interventions or the overall autonomy depends on (i)~the node's energy consumption per day, (ii)~the energy storage technology and (iii)~the selected voltage conversion~\cite{callebaut2021art}. We model the energy consumption of a node equipped with a sensor, an \gls{lpwan} technology and a battery. Based on that model, the energy provisioning requirement is determined.



Eq.~\ref{eq:chgperinv} defines the maximum charged energy per intervention. This depends on the charge rate (CR) [1/h], the total battery capacity ($E_{\text{Battery}}$) [J] and the charge time ($T_{\text{Charged}}$) [s/intervention]. We assume a linear relationship between $T_{\text{Charged}}$ and $E_{\text{charged}}$. In practice, the energy-intake will depend on the \acrlong{soc}. The \gls{soc} determines whether the battery is charged in \gls{cv} or \gls{cc} mode, resulting in a non linear behavior in practice. 
We assume that  the battery will be almost fully depleted before recharging, and hence the majority of the charge time will follow the linear behavior.

\begin{equation}
    E_{\text{Charged}} = E_{\text{Battery}} \cdot \frac{\text{CR}}{3600} \cdot T_{\text{Charged}}
    \label{eq:chgperinv}
\end{equation}

The required energy transfer per intervention is given in Eq.~\ref{eq:ereq} and takes into account the energy-efficiency of the \gls{wpt}. Multiple components contribute to the overall energy losses per day ($E_{\text{Losses}}$) [J/day], which consists of the sum of self discharge losses ($E_{\text{SD}}$), conversion losses ($E_{\text{Conv}}$) and leakage losses ($E_{\text{Leak}}$).
The minimum required energy, to bridge the days between two interventions, becomes
\begin{equation}
    E_{\text{Req,min}} = (E_{\text{Consumed}} + E_{\text{Losses}}) \cdot \frac{365}{n}\ ,
    \label{eq:ereq}
\end{equation}
with $n$ the number of interventions per year. Hence, in order to ensure a continuous operation $E_{\text{Battery}}$ should be higher than $E_{\text{Req,min}}$. If $E_{\text{Charged}} > E_{\text{Req,min}}$, theoretically, an infinite autonomy is ensured. In this specific case, the charge time could be adapted or the interventions per year diminished to save energy of the \gls{uav}. However, if $E_{\text{Charged}} < E_{\text{Req,min}}$, the autonomy $T_{\text{Aut}}$ [Days] is limited and could be calculated with Eq.~\ref{eq:autonomy}. 

\begin{equation}
    T_{\text{Aut}} = \frac{365 \cdot E_{\text{Battery}}}{365 \cdot (E_{\text{Consumed}} + E_{\text{Losses}}) - E_{\text{Charged}} \cdot n}
    \label{eq:autonomy}
\end{equation}

The condition  presented in Eq.~\ref{eq:cond} must be met to achieve a virtually unlimited autonomy. It takes into account the minimal required energy to keep the node running between two interventions. 
In addition, the battery capcity is limited. As a consequence, the stored energy per intervention $E_{\text{charged}}$, constrained by $E_{\text{battery}}$, needs to be higher than the required energy to guarantee the operation of the node $E_{\text{Req,min}}$. 




\begin{equation}\label{eq:cond}
    E_{\text{Battery}} > \max\left( \frac{E_{\text{Req,min}} \cdot 3600}{\text{CR} \cdot T_{\text{Charged}}},E_{\text{Req,min}}\right)
\end{equation}


Eq.~\ref{eq:cond} allows the designer to select the battery capacity as low as possible, resulting in a more environmentally friendly device. To do so, one can increase n, CR or $T_{Charged}$, while maintaining a certain autonomy.

\section{Use case-based assessment}

We evaluate both high-energy and low-energy consuming \gls{iot} use cases in view of designing a solution for contactless energy provisioning by \glspl{uav}. 
One low-power application measures temperature difference at the trees to determine their health~\cite{thoen2019deployable}. Another application uses a volatile gas sensor (Bosch BME680) to monitor the air quality. These gas sensors need to be heated to a target temperature of \SI{320}{\degree C}, taking around \SI{92}{s}, yielding a high energy-penalty per measurement. This low-power use case will be further denoted as the \textit{Tree node} application, while the more energy-demanding use case will be called the \textit{Gas node} application. 

In fair comparison, the same wireless transceiver, \gls{mcu} and batteries are used in the evaluation, as presented in Table~\ref{tab:usecasesenergycons}. 
The \gls{mcu} sleep power is \SI{0.025}{mW}. During operation, the node wakes-up, reads out the sensors and transmits the data via LoRaWAN to the cloud. We assume for this transmission a spreading factor~of~12 with a bandwidth of \SI{125}{kHz} and a payload size of \SI{20}{bytes}~\cite{thoen2019deployable}. This yields a power consumption of \SI{111.15}{mW} over a time duration of \SI{1810}{ms}. To consider a realistic example, we assume a data transmission every 15~minutes. The energy consumption per measurement, including wireless transmission and sensor reading, amounts to \SI{0.012}{J}~\cite{thoen2019deployable} for the low-power sensor and \SI{2.153}{J} for the high-power sensor. The consumed energy and duration for both nodes is summarized in Table~\ref{tab:usecasesenergycons}. In total, the \textit{Tree node} and \textit{Gas node} consume, respectively, \SI{22.7}{\joule} and \SI{227.9}{\joule} per day. 
\begin{table}[btp]
    \vspace{4pt}
    \centering
    \caption{Energy consumption for the two use cases}
    \label{tab:usecasesenergycons}
    \begin{tabular}{lrrrr}
    \toprule
    Energy Consumption & \multicolumn{2}{c}{Tree node}  & \multicolumn{2}{c}{Gas node}\\
    \cmidrule(lr){2-3} \cmidrule(lr){4-5}
    & mW & s & mW & s\\
    \midrule
    MCU sleep          & \num{0.025} & -    & \num{0.025} & -      \\
    LoRaWAN (SF12)       & \num{111.1} & 1.81    & \num{111.1} & 1.81     \\
    Sensor             & \num{65.7} & 0.19     & \num{23.4} & 92.00    \\
    \bottomrule
    \end{tabular}
    \vspace{10pt}
\end{table}

In the following, both non-rechargeable and rechargeable batteries are evaluated as an energy source for these use cases.

\subsubsection{Non-rechargeable solution}

The power source of the \textit{Tree node} system exists of three in series-connected alkaline batteries, with a usable \SI{21}{\kilo\joule} of energy. An autonomy of \SI{2.5}{year} could be theoretically achieved~\cite{thoen2019deployable}. The \textit{Gas node}, with a similar autonomy, needs a battery pack of \SI{208}{\kilo\joule} or, equivalently, \num{30}~alkaline cells on the node side. Here, you can already state that providing this amount of stored energy is infeasible and not the least, not environmental friendly.


\subsubsection{Rechargeable solution}
\begin{figure}[!hbt]
    \centering
    \input{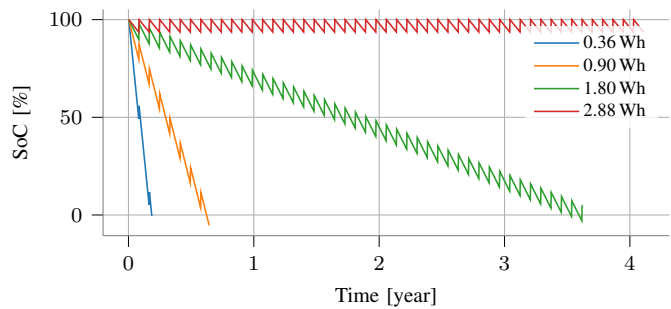}
    \caption{The \acrlong{soc} degradation depends on the selected battery. A higher capacity will automatically give higher autonomy.}
    \label{fig:soc_lifetime}
\end{figure}

To extend the autonomy, a rechargeable energy storage is required. Furthermore, the capacity of the battery can be drastically lowered, thereby requiring less materials for the same or even longer autonomy. 
In the following analysis, we assume a \textit{Tree node} consuming a fixed amount of energy every day. We demonstrate the importance of choosing the best fit type of storage technology. For demonstration, we start with an \gls{lco} rechargeable battery with a nominal cell voltage of \SI{3.6}{\volt} and a variable capacity. The \gls{soc} degradation of the battery over time is shown in Fig.~\ref{fig:soc_lifetime}. 
Assuming 12~interventions per year, a charge rate restriction of 1C and a charge time of \SI{5}{\minute}, a theoretical infinite autonomy is obtained. 
Choosing the battery capacity too small will quickly lead to a depleted node. 
Continuing on this example, the optimal battery capacity for powering the \textit{Tree node} can be determined and will be situated between \SI{1.80}{Wh} and \SI{2.88}{Wh}. 

\begin{figure}[!hbt]
    \centering
    \input{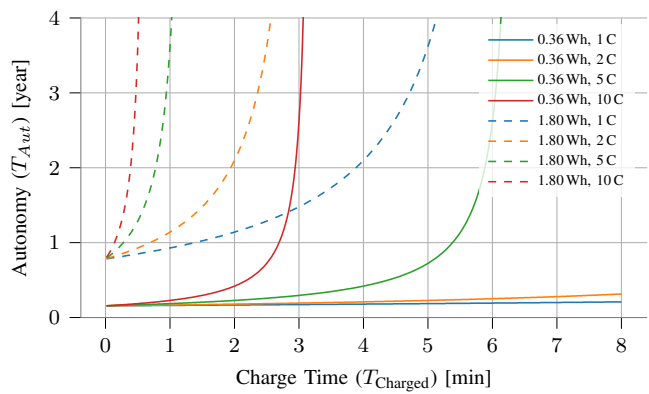}
    \caption{Autonomy in function of charge time, battery capacity and battery C-rate. Number of intervention is fixed to 12.}
    \label{fig:aut_chg_time}
\end{figure}

Secondly, the autonomy is related to the charge time. Increasing $T_{\text{Charged}}$ results in more energy stored during each intervention, as depicted in Fig~\ref{fig:aut_chg_time}. Two battery capacities (\SI{0.36}{Wh} and \SI{1.80}{Wh}) for different battery technologies with corresponding charge rates (\SI{1}{C}, \SI{2}{C}, \SI{5}{C} and \SI{10}{C}) are considered. The required charge time decreases, when a battery technology with higher charge rate is selected. 

\begin{figure}[!hbt]
    \centering
\begin{tikzpicture}

\definecolor{color0}{rgb}{0.12156862745098,0.466666666666667,0.705882352941177}
\definecolor{color1}{rgb}{1,0.498039215686275,0.0549019607843137}
\definecolor{color2}{rgb}{0.172549019607843,0.627450980392157,0.172549019607843}
\definecolor{color3}{rgb}{0.83921568627451,0.152941176470588,0.156862745098039}

\begin{axis}[
tick label style = {font = \footnotesize},
scale only axis,
width = 0.85\linewidth,
height = 0.45\linewidth,
xmin=0, xmax=25.5,
ymin=0, ymax=72.0,
axis line style={white!50.19607843137255!black},
axis lines* = {left},
tick align=outside,
legend cell align={left},
legend style={fill opacity=0.8, draw opacity=1, text opacity=1, draw=white!80!black, font=\footnotesize},
tick pos=left,
x grid style={white!69.01960784313725!black},
xtick style={color=black},
y grid style={white!69.01960784313725!black},
ylabel={\footnotesize $E_\text{battery}$ [\si{\kilo\joule}]},
xlabel={\footnotesize Number of interventions per year ($n$)},
ytick style={color=black},
legend columns=2,
legend pos=north west,
ymajorgrids,
xmajorgrids,
legend style={draw=none, nodes={scale=0.7, transform shape}},
tick label style = {font = \footnotesize},
]

\addplot[domain=2:24, no markers, color0, thick] {(22*365*3600)/(300*(x*1000))};
\addlegendentry{Tree node - \SI{1}{C}}

\addplot[domain=2:24, no markers, color1, thick] {(210*365*3600)/(300*(x*1000))};
\addlegendentry{Gas node - \SI{1}{C}}

\addplot[domain=2:24, no markers, color0, dashed, thick] {(22*365*3600)/(10*300*(x*1000))};
\addlegendentry{Tree node - \SI{10}{C}}

\addplot[domain=2:24, no markers, color1, dashed, thick] {(210*365*3600)/(10*300*(x*1000))};
\addlegendentry{Gas node - \SI{10}{C}}

\end{axis}

\end{tikzpicture}%
    \caption{The minimal required energy for the Tree node and Gas node in combination with two types of batteries.}
    \label{fig:E_batt_vs_interventions}
\end{figure}
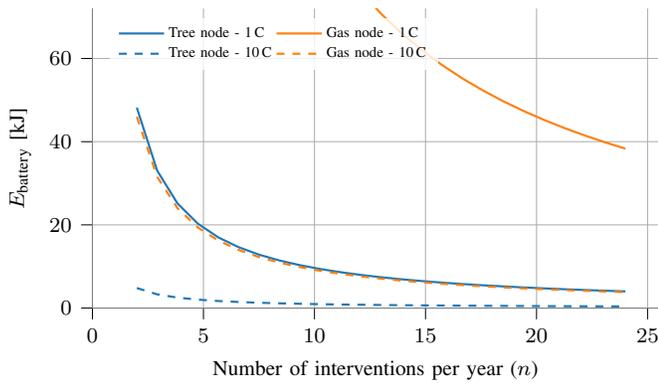

Finally, the required battery capacity with respect to the number of interventions is assessed for the two use cases and is depicted in Fig.~\ref{fig:E_batt_vs_interventions}. We distinguish two lithium based batteries: an \gls{lco} and \gls{lto} battery with a charge rate of respectively \SI{1}{C} and \SI{10}{C}, giving in total four curves. Here, the charge time during an intervention is assumed to be restricted to \SI{5}{\minute}. Assuming that the interventions per year are predefined, the selection of a \SI{10}{C} capable battery allows the designer to reduce the battery capacity significantly compared with a \SI{1}{C} variant. Otherwise, for the same amount of battery capacity, more interventions are required, when choosing a \SI{1}{C} battery above a \SI{10}{C} variant.
To reduce the battery capacity and electronic waste, \gls{lto} technology can be considered as energy storage for \gls{uav} rechargeable nodes, as it provides a higher c-rate than alternatives. Further, the higher number of charge cycles, wider operating temperature and higher chemical stability makes them attractive to use in the considered scenarios. A disadvantage of this technology is the lower volumetric energy density and low nominal voltage compared to an \gls{lco} battery~\cite{han2014cycle}. Although, with recent developments, extremely efficient \gls{smps} with low quiescent currents could convert the low nominal cell voltage to a fixed output voltage.
\section{Conclusion}
\label{conclusion}
We presented a \gls{uav}-based system to increase the autonomy of remote and difficult accessible \gls{iot} nodes. Multiple advantages are associated with this concept such as the eco-friendly benefits and proactive maintenance. Although this concept comes with challenges related to both the localization and the charging of the devices, the approach can outperform alternative solutions such as manual replacement/recharging or long-distance RF-based charging, by orders of magnitude in terms of energy efficiency. Adequate technologies need to be selected for the energy storage and the energy transfer respectively. 
We proposed a low complexity model to estimate the required battery capacity based on a number of predefined parameters. 
We showed that the capacity can be reduced by increasing the charging speed, thus it should be considered to use energy storage technologies with a high c-rate or a low internal resistance. In our future work, we are optimizing an actual design, whereby the \gls{uav} will recharge itself based on a sustainable source. 

\bibliographystyle{IEEEtranN}
{\footnotesize
\bibliography{bronnen}}

\end{document}